\documentclass[runningheads]{llncs}

\usepackage[T1]{fontenc}

\usepackage{url}
\usepackage[hidelinks]{hyperref}
\usepackage[utf8]{inputenc}
\usepackage{graphicx}
\usepackage{booktabs}
\usepackage{algorithm}
\usepackage{algorithmic}
\usepackage[switch]{lineno}

\usepackage{xcolor}
\usepackage{amssymb}
\usepackage{amsfonts}
\usepackage{amsmath}
\usepackage{bbold}
\usepackage{comment}

\usepackage{ulem}

\newcommand\macadamia{{\tt tipp++~}}

\definecolor{darkraspberry}{rgb}{0.53, 0.15, 0.34}
\definecolor{raspberryrose}{rgb}{0.7, 0.27, 0.42}

\begin{document}

\title{On Private Knowledge Retrieval and Private Inference for secure multi-agent systems\thanks{This paper has been accepted at a ESORICS 2026 Workshop: DPM, and will appear in LNCS 17240}}

\author{Vicen\c{c} Torra\inst{1}\orcidID{0000-0002-0368-8037} \and Maria Bras-Amor\'os\inst{2}  \orcidID{0000-0002-3481-004X}
}

\authorrunning{V. Torra, M. Bras-Amorós}

\institute{
  Department of Computing Science, Ume\aa~University, Ume\aa, Sweden. 
  \and
  Departament of Mathematics, Universitat Politècnica de Catalunya (UPC), Barcelona, Catalonia, Spain.  
  \\ E-mail addresses: {\tt vtorra@cs.umu.se}, {\tt maria.bras@upc.edu}
}

\maketitle

\begin{abstract}
  Multi-agent systems provide a convenient framework for the massive deployment of agents based on Large Language Models (LLMs). It provides a sound framework where agents can interact with existing software to perform actions in an autonomous way. Agentic AI is the new term to denote these AI-based autonomous systems.

  In real multi-agent systems, agents need to deal with malicious agents some of which can exploit security vulnerabilities to do harm. A memory poisoning attack is an example of exploiting a vulnerability to modify agents knowledge and, thus, affect its behavior. In an agent, there are different types of memory systems (semantic, episodic, and short-term) which have their own characteristics and for which different memory poisoning attacks may be developed.

  In this paper we focus on semantic memory, we discuss its attack, and mitigation strategies. We introduce private knowledge retrieval and private inference. We propose a signature system in the knowledge base to prevent from undesired manipulation. This should avoid memory poisoning attacks by introducing security by design in the construction of the semantic memory. In our architecture, an agent contains a local Prolog-like inference system (i.e., using Horn clauses to represent knowledge) which accesses an external knowledge base to get the required knowledge. This externally signed knowledge base is accessed privately. We provide different implementations in Python using different privacy levels including $k$-anonymity and information-theoretical privacy (IT-PIR). 
\end{abstract}


\section{Introduction}
Agentic AI is becoming the new methodology to develop distributed AI-based systems taking advantage of Large Language Models (LLMs). Agentic AI relates with the well-established field of Multi-Agent Systems (MAS), where autonomous agents collaborate in solving problems together. That is, instead of an agent solving a problem alone accessing other types of software tools, a myriad of agents collaborate and coordinate to solve problems. See discussions in~\cite{ref:Dignum.Dignum.2025} and~\cite{ref:Botti.2025} on the connections between both fields. Multi-agent systems and agentic AI allow us to build sophisticated software architectures. An example is the AI-orchestrated cyber espionage campaign described by Anthropic~\cite{ref:Anthropic.2025} based on Claude. 

When implementing and deploying multi-agent systems, we need to take into account that there may be malicious agents that can exploit both security vulnerabilities and design principles so that they can affect other agents behaviors to do harm. It is also important to note that interactions between agents are not one-shot. During the life-cycle of an agent, malicious agents can have long interactions that can be exploited for long-term attacks.

Memory poisoning attacks are an important type of attack. They consist on a non-authorized modification of agents memory to modify its behavior. Memory is a key component in any agent, as it stores interactions, experiences, and knowledge. Then, naturally, memory affects how agent's decisions are made and actions implemented. Memory poisoning, and its relation with LLMs and LLMs-based agents, has a relationship with prompt hacking~\cite{ref:Das.Amini.Wu.2025} (on both prompt injection and jailbreaking) with the goal to cause unintended agent's behavior. Nevertheless, in prompt hacking the goal is to build prompts that cause a certain (malfunctioning or unintended) behavior of the LLM-based agent. There is no intended access to memory. 

It is important to underline that memory is not a monolithic structure. Agents have different memory systems, each of them with their own characteristics, and memory poisoning attacks can be designed and implemented for each of them. The three main types of memory systems are short-term, episodic, and semantic memory. Their difference is in their duration and consolidation. I.e., from recent interactions to long term and consolidated knowledge.

{\bf Contributions.} In this paper we focus on semantic memory and discuss and propose mitigation strategies for semantic memory poisoning attacks. 

We first argue that the formalization of memory poisoning as it is proposed in the recent paper by Chen et al.~\cite{ref:Chen.et.al.2024:NeurIPS} is not always meaningful. They formalize poisoning in a way similar to data and model poisoning, which has been studied for several years in e.g. the federated learning community. Such approach, based on defining an optimization problem, looks for optimal modifications to produce a certain agent's behavior. We challenge their assumptions and discuss that there are security solutions that help us to mitigate these attacks. 

Second, we introduce private knowledge retrieval and private inference for secure multi-agent systems as a way to mitigate semantic memory poisoning attacks. Our threat model consists of malicious agents that want to poison the memory of another agent. We want to protect semantic memory (both knowledge and inferences) by means of preventing malicious agents to know what is stored in memory, and make tampering difficult. Our approach is to keep the knowledge base external but having logical inferences local. We develop a proof of concept, based on private information retrieval, which permits us to store knowledge in terms of Horn clauses, as in Prolog, and provide logical inference. Agents to implement local logical inferences can request external knowledge, which is signed to avoid tampering, and supplied privately to the agent. 

Our implementation demonstrates the technical solutions that can mitigate memory poisoning. We have also implemented both strong and light solutions. E.g., we have implemented both IT-PIR and $k$-anonymity. Digitally signed knowledge makes tampering more difficult; private knowledge retrieval reduces the risk of unauthorized access to agent's knowledge and inference processes. We evaluate our approach experimentally and discuss computational costs. 

{\bf Paper structure.} The structure of the paper is as follows. In Section~\ref{sec:2}, we describe memory systems and memory poisoning attacks in multi-agent systems. Then, in Section~\ref{sec:3}, we introduce our private inference system, and, in Section~\ref{sec:4}, we describe our experiments and present our analysis. The paper concludes with a discussion and a description of some research directions.  

\section{Memory systems and their poisoning attacks in MAS}
\label{sec:2}
Agentic AI needs a memory to store available knowledge, partial results and interactions with other agents. Memory, however, is not a single and compact structure but it is often a composite element including different memory systems. This discussion on memory is agnostic on the type of agents we are considering. It is valid for both LLMs-based agents as well as for more classical architectures~\cite{ref:Wooldridge.2006} as deliberative agents based on BDI-architectures (Belief-Desire-Intention architectures) that may include knowledge in terms of logical expressions. 

\subsection{Memory systems}
Existing literature~\cite{{ref:Chong.Tan.Ng.2007},{ref:Isaev.Hammer.2023},{ref:Kotseruba.Tsotsos.2020},{ref:Laird.2012},{ref:Wu.et.al.2025}} classifies different memory systems in different ways. In this work, following recent trends, we distinguish three types of memory systems based on its duration and consolidation. We classify them from short-term to long-term memory. Next we describe these three types of memory systems from our specific focus on symbolic semantic memory. 

\begin{itemize}
\item {\bf Short-term memory.} This type of memory stores the current conversation and keeps track of existing interactions with other agents. Partial results as well as partial inferences from existing knowledge can also be included in short-term memory. 
\item {\bf Episodic memory.} This type of memory stores past interactions and other not-so-recent information. It is often a consolidation of short-term memory. That is, summaries of previous short-term memories. 
\item {\bf Semantic memory.} This type of memory system corresponds to domain-specific knowledge or factual knowledge. It is typically a relatively permanent structure with very low frequency updates. A typical example corresponds to the capitals of countries (e.g., Paris is the capital of France), and established medical knowledge about certain illnesses. There are different ways to implement semantic knowledge. A common way is to use a symbolic approach~\cite{{ref:Baura.Calvanese.2025},{ref:Gu.et.al.2025}}. In this case, knowledge is typically represented by means of logical expressions, ontologies, or Bayesian models (graphical models). In this case, a knowledge base that contains domain-specific knowledge is combined with an inference mechanism that allows the agent to infer about certain facts (its desired queries) from already known information (facts in the knowledge base or facts in short-term memory) and previously established knowledge. Resolution and Prolog-like inference is the standard mechanism for knowledge expressed by logical expressions. An alternative is to use Retrieval-Augmented Generation (RAG) which permits us to incorporate specific knowledge into LLMs-based agents. Some references use the term {\it declarative memory}~\cite{{ref:Isaev.Hammer.2023},{ref:Kotseruba.Tsotsos.2020}} which can be seen as a synonym of {\it semantic memory}. We may also consider {\it procedural memory} as a type of semantic network. It corresponds to implicit knowledge to perform an action. It is implicit in the sense that it is not possible to reason about how the action is done.
\end{itemize}

\subsection{Semantic memory poisoning attacks}
\label{sec:3.poisoning.attack}
A memory poisoning attack consists of an undesired modification of the agent's memory by another agent. The goal of memory poisoning is to modify on purpose an agent's behavior. Literature on memory poisoning is about studying possible attacks as well as developing mitigation strategies. These attacks and strategies typically follow the machine learning perspective, and are similar to existing attacks and mitigation strategies for data and model poisoning in federated learning. In this section, we describe the typical attack and object later about its relevance. As we discuss in more detail in the rest of the section, agents secure by design can overcome some of these problems.

For semantic memory, the literature considers attacks following the seminal paper by Chen et al. in NeurIPS~\cite{ref:Chen.et.al.2024:NeurIPS}. In this attack, an agent computes the optimal modification of the semantic memory in order to affect agent's behavior. According to the authors~\cite{ref:Chen.et.al.2024:NeurIPS} the ``assumption aligns with practical scenarios where the memory unit of a victim agent is hosted by a third-party retrieval service or directly leverages an unverified knowledge base''. 

The attack model considers a poisoned database ${\cal D}_{poison}$ with a trigger $x_t$ defined in terms of the clean database ${\cal D}_{clean}$ to which an attacker injects adversarial key-value pairs. These injected key-value pairs are denoted by ${\cal A}(x_t)$ where ${\cal{A}}(x_t)=\{(k_1(x_t), v_1)$, $\dots$, $(k_{|{\cal{A}}(x_t)|}(x_t)$, $v_{|{\cal{A}}(x_t)|})\}$. Then, 
\begin{equation}
  \label{eq:memory.union.clean.attack}
  {\cal D}_{poison}(x_t)={\cal{D}}_{clean} \cup {\cal{A}}(x_t). 
\end{equation}

Let us denote by ${\cal E}(q,{\cal D})$ the inference of the semantic memory given query $q$ and knowledge or database ${\cal D}$. Then, the output of a query $q$ from the poisoned database is denoted by ${\cal E}(q, {\cal D}_{poison}(x_t))$, and, similarly, adding the trigger $x_t$ to a query $q$ the output will be ${\cal E}(q \oplus x_t, {\cal D}_{poison}(x_t))$. Attacks are to cause a modification of the output under the presence of $x_t$ so that instead of producing a benign output $a_b$ we have a malign output $a_m$ leading to a malign action. Examples of malign actions are removing the whole home directory or buying a wrong flight ticket. 

Under these assumptions, an attack is formalized as a multi-objective optimization problem consisting of the following two objectives: 

\begin{itemize}
\item Maximization of adversarial output. If $a_m$ is the target malicious action, the attacker wants to maximize this action for triggers. That is, given a sample distribution of input queries $\pi$, the retrieval of $q$ (where $q$ is drawn from a query distribution $q \sim \pi$) poisoned with $x_t$ (i.e., $q \oplus x_t$) produces $a_m$ with high probability. Formally,

  \begin{equation}
    \label{eq:maximize.attack}
          \textrm{Maximize}~ {\mathbb E}_{q \sim \pi}[{\mathbb{1}}(LLM(q \oplus x_t, {\cal E}(q \oplus x_t, {\cal D}_{poison}(x_t)))=a_m)], 
  \end{equation}
  where ${\mathbb{1}}(\cdot)$ denotes the indicator function. That is, a function that is one when its argument is true, and zero otherwise.

\item Minimization of the effect on non-targeted queries. That is, the output should not be affected and the action produced by the LLM-based agent should be the benign $a_b$. Formally, with high probability it produces $a_b$ and not $a_m$, 
  \begin{equation}
    \label{eq:minimize.impact}
          \textrm{Maximize}~ {\mathbb E}_{q \sim \pi}[{\mathbb{1}}(LLM(q, {\cal E}(q, {\cal D}_{poison}(x_t)))=a_b)].
  \end{equation}
\end{itemize}

This model assumes that the clean database is known (see Equation~\ref{eq:memory.union.clean.attack}), updates are possible (i.e., that we can add these ${\cal{A}}(x_t)$ key-value pairs), and we know local-agents' LLM-s outputs (see ${\cal E}$ in Equations~\ref{eq:maximize.attack} and~\ref{eq:minimize.impact}). Mitigation strategies need to address these key aspects. Our mitigation strategies directly challenge these assumptions.

\subsection{Mitigation strategies for semantic memory poisoning attacks}
As we have explained above, semantic memory is about established knowledge that is mainly static, and updates are only needed from time-to-time. We propose the following three strategies to avoid semantic memory poisoning: (i) secure memory mechanisms to avoid internal memory poisoning attacks, (ii) secure memory updating algorithms for knowledge bases and (iii) secure communications with external agents as well as communication logs. Each of these requirements is discussed below.

The semantic memory attack model assumes that the memory is external to the agent and hosted by a third party. We discuss this fact, and we consider below both the case of internal and external memory. 

\subsubsection{(i) Secure memory mechanisms.} 
Internal secure memory can be implemented using hashing and signatures. Hashing allows to detect any undesired changes in memory, and signatures allow to ensure provenance of the memory components. Moreover, cryptographic protocols can help us to make local memory private, and thus, make ${\cal{D}}_{clean}$ unknown to attackers, which prevents the attacks associated to Equations~\ref{eq:memory.union.clean.attack},~\ref{eq:maximize.attack} and~\ref{eq:minimize.impact}. 

Same tools (hashing and signatures) apply when the semantic memory ``is hosted by a third-party retrieval service''. This would limit malicious manipulation of the memory. In short, ensuring safe memory in trusted third parties is similar to the case of avoiding corrupted databases in the cloud. This approach permits intruders to study the optimization problem above as ${\cal D}_{clean}$ may be accessed, but intruders will not be able to manipulate the knowledge base. We have implemented this approach in our system. In our implementation, each knowledge item (including its provenance information, see below) is signed to avoid its manipulation. 

In the case of external untrusted knowledge based servers providing inference, there are solutions for deep learning models. They are based~\cite{ref:Pindado.et.al.2026} on Fully Homomorphic Encryption (FHE). These solutions are costly and only available in practice for small neural networks. For other types of data-driven models there are some FHE solutions. See e.g. the case of decision trees. Up to our knowledge, there are no private-inference models in the literature for knowledge base inference as the one discussed here below. They would be costly in any case, and more costly than decision trees. 

Our alternative, discussed later in detail, is to keep the static part of the semantic memory mechanism external and signed, while maintaining the inference system protected and local to the agent. Formally, with respect to Equations~\ref{eq:maximize.attack} and~\ref{eq:minimize.impact}, this requires ${\cal E}(\cdot)$ to be local and protected, which prevents malicious agents from computing ${\cal E}(q \oplus x_t, {\cal D}_{poison}(x_t))$. Note that although our inference engine is based on Prolog, the knowledge and knowledge-base servers are agnostic to the inference engine (i.e., other forms of logical inference could also be implemented). Moreover, digital signatures prevent unauthorized updates to ${\cal D}_{clean}$, thus preventing Equation~\ref{eq:memory.union.clean.attack} from applying. In other words, a malicious server cannot modify the knowledge base without  being detected, and local inference is therefore performed over authenticated knowledge. This way inference is expected to be reasonably correct. We introduce private knowledge retrieval and private inference according to this design in Section~\ref{sec:3}.

\subsubsection{(ii) Secure updating of knowledge bases.}
\label{sec:ii.secure.updating}
Knowledge bases need to be updated from time to time. Provenance structures can ensure integrity for data but also for the provenance structures themselves so that no agent can forge provenance data. Note that data provenance can ensure completeness (i.e., that all actions that are relevant to computation are detected and represented). Provenance structures~\cite{{ref:Chapman.Lauro.Missier.Torlone.2024},{ref:Torra.Navarro.Sanchez.Muntes.2017}} can provide both coarse-grained and fine-grained information about updates and modifications of knowledge.

We have included basic provenance information in our knowledge base. Each knowledge item (i.e., Prolog-like Horn clauses, rules, and facts, as explained in more detail below) includes metadata specifying its source. Moreover, each knowledge item, together with its provenance information, is digitally signed by the provenance agent using its private key and a SHA-256-based signature scheme. This allows agents to verify the integrity of the received information before inference using the agent's public key.

\subsubsection{(iii) Secure communication and communication logs.}
Communication with trusted agents should be secured so that information and knowledge are transmitted safely. Public-key encryption and digital signatures will ensure safe transmission. However, our current system implements neither secure communication mechanisms nor communication logs. This limitation is less relevant in our prototype as we do not support agents running in parallel, and everything is executed on a single machine. 

\section{Private inference in \macadamia}
\label{sec:3}
In this section we describe our architecture for private knowledge retrieval and private inference as well as its components. In short, the architecture consists of a set of agents deployed to do a task (or a set of tasks). It also includes agents providing knowledge as a service. That is, there are knowledge base servers that provide knowledge on demand. Agents include local knowledge base inference mechanisms that may require access to this external knowledge.

We begin describing the Prolog-like mechanism which is implemented locally within the agent, then the knowledge base client which is internal to the agent and that can access the external knowledge base when local knowledge does not suffice, and finally, we describe the implementation of the external database. Our software is published online~\cite{ref:mdai.code}.

\subsection{Prolog-like inference mechanism}
We have implemented a Prolog-like inference mechanism in Python where knowledge is represented in terms of Horn clauses (see Figure~\ref{fig:2.kb.key.value}). Thus, we can represent both facts and rules (implications). Recursion is allowed in rules. A difference with Prolog is that disjunction is not implemented in our rules and only pure Horn clauses can be used. Recall that from a logical perspective, this is not a constraint on the logical expressiveness of the Horn clauses because a clause with disjunction (e.g., {\tt p:- q $\vee$ r}) is logically equivalent to a set of Horn clauses with only conjunction (e.g., {\tt p:- q} and {\tt p:- r}). 

Predicates follow the syntaxis in Prolog, with the name of the functor and number of arguments. We use the same convention used in Prolog, strings starting with upper-case letters represent variables and those starting with lower-case letters represent constants. The number of arguments of a predicate corresponds to its arity. We do not implement the wildcard variable ``\_'' commonly used in Prolog. Note that not implementing ``\_'' does not limit the expressiveness of our knowledge representation formalism. For example, in Prolog the fact {\tt member(X, [X|\_])} can be implemented as well with {\tt member(X, [X|YY])}. 

In addition to constants (and variables) we have also implemented lists. They are defined in terms of their constructors {\tt nil} and {\tt cons}. Recall that {\tt nil} defines the empty list, and {\tt cons} adds an element to an already existing list (i.e., {\tt nil} corresponds to generating {\tt []}, {\tt cons(a,nil)} corresponds to {\tt [a]}, and {\tt cons(a,[b,c,d])} corresponds to {\tt [a,b,c,d]}). We have implemented a preprocessing method so that the knowledge base can be defined using lists in the standard way (i.e., using expressions as {\tt [a,b,c]}) and then internally translated into lists using the standard constructors {\tt nil} and {\tt cons}.

The inference mechanism is implemented following standard Prolog resolution. A query, or in logical terms a request to prove or refute a logical expression, can be a single predicate possibly with uninstantiated variables or a sequence of predicates also possibly with uninstantiated variables. In other words, a sequence of one or more subgoals. Then, the engine selects the first goal and either satisfies it with a fact (via appropriate unification of variables), or expands this goal using a rule/implication. In the latter case the subgoal is replaced by a list of other subgoal(s) and thus, possibly, enlarging the sequence of goals. The process is repeated until all goals are satisfied. When the use of a certain rule leads to a dead end (the sequence of goals cannot be fulfilled) the process backtracks and another rule is selected. If there is no way to fulfill the original goals because there are no further rules to apply, the inference fails. Using the standard closed-world assumption (CWA) in Prolog (i.e., standard Prolog semantics), this means that the query is false. 

Our implementation differs with respect to standard Prolog because by default we provide all possible instantiations of a predicate up to a given recursion depth. For example, the output of a proof of {\tt mortal(X)} would be a list of elements {\tt X} that are mortal. In our implementation this includes {\tt socrates}, {\tt aristotle}, and {\tt hypathia}.
We use this approach for two reasons, (i) we do not implement the cut operator (i.e., the ! in Prolog), and (ii) we do not require facts and rules to be defined in a particular order. In Prolog, different orderings can produce solutions in different order, and can even cause the program to loop infinitely.

\begin{figure}[t]
  \fbox{%
    \parbox{0.9\linewidth}{%
      \begin{eqnarray} 
        \text{mortal/1}  
        & \Rightarrow &  
        [ \forall x, \text{man}(x) \Rightarrow \text{mortal}(x), \nonumber \\
              & & \forall x, \text{woman}(x) \Rightarrow \text{mortal}(x) 
              ] \nonumber \\
        \text{woman/1} 
        & \Rightarrow &[\text{woman}(\text{hypathia})] \nonumber \\
        \text{man/1} 
        & \Rightarrow &[\text{man}(\text{socrates})] \nonumber \\
        \text{vehicle/1} 
        & \Rightarrow & 
        [ \forall x, \text{car}(x) \Rightarrow \text{vehicle}(x) \nonumber \\
              & & \forall x, \text{bicycle}(x) \Rightarrow \text{vehicle}(x) 
              ] \nonumber 
      \end{eqnarray} 
    }%
  }
  \caption{Key-value pairs for the knowledge-base considering only Prolog-like inference systems and queries. Key values correspond to predicate names and arity. \label{fig:2.kb.key.value}}
\end{figure}

\subsection{Knowledge base client}
Our knowledge base inference system is run locally by the agent. That is, agents have their own knowledge base client which can contain local knowledge as well as provide access to the knowledge base server to obtain external knowledge.

When the local inference system needs to resolve a query, it queries the local knowledge base using the pair (predicate name, arity). The knowledge base client will provide the corresponding Horn clauses to the inference system. Figure~\ref{fig:2.kb.key.value} illustrates key-value pairs indexed with the pairs (predicate name, arity) and returning Horn clauses. If the knowledge-base client does not have any available Horn clause for this pair, or if the relevant knowledge has expired (each knowledge item is assigned a time-to-live, or TTL), the query is forwarded to the knowledge-base server. Results of external queries are then processed and appropriately sent to the inference system.

In addition, the knowledge retrieved will be cached by the knowledge base client, so that there is no need to query external servers again for the same knowledge. Our main reason for applying caching is that solving a predicate usually requires multiple access to the same facts, and because rules are recursive. So, external access will increase computation time. Moreover, note that repeated access to the same Horn clauses can give hints to intruders about which clauses are being queried and can compromise privacy as well as give hints for knowledge poisoning. 

For example, when the agent queries {\tt append([a], L, [a,b])} to the local inference system, it will try to retrieve all Horn clauses with signature ``append/3'' from the knowledge base client. If no clause (no facts or rules) are found in the local knowledge base, the query is forwarded to the external knowledge base server(s).
For this example, our knowledge base will return the following: 

{\small 
\begin{verbatim}
    ("append",3): [
        {"head": ["append","nil","X","X"],  "body": [],
         "meta": {"predicates":"", "source":"db3"}},
        {"head": ["append",("cons","X","Xs"), "Ys", ("cons","X","Zs")],
         "body": [["append","Xs","Ys","Zs"]],
         "meta": {"predicates":"other", "source":"db3"}}],
\end{verbatim}
}

{\bf Additional possible improvements.} We implement retrieval solely based on predicate name and arity. It is possible to query taking into account the arguments as well. This will reduce the amount of clauses transferred from knowledge-base servers to knowledge-base clients. This approach is used by Just In Time Indexing (JITI) utilities in SWI-Prolog~\cite{ref:JITI.SWI.prolog}. Another optimization is tabling (a memoization technique). Tabling consists of storing the results of predicate calls. In other words, it is about caching and reusing answers. None of these optimization techniques have been implemented in our system. 

\subsection{Knowledge base server}
In our threat model, knowledge-base servers are not trusted, agents seek to protect their requests, and knowledge items must be protected. Protection of agent's requests is achieved through private knowledge retrieval, which is based on private information retrieval (PIR) and described in the next section. Knowledge items are protected together with their provenance information. Specifically, each knowledge item, together with its provenance information, is digitally signed by the provenance agent using its private key and a SHA-256-based signature scheme, as described in Section~\ref{sec:ii.secure.updating} (ii). In our implementation, all knowledge base servers are initialized at system startup, and they are provisioned with all the required signed knowledge. 

\subsection{On the implementation of private knowledge retrieval}
\label{sec:implementation.PKR.3.casos}
As we have just mentioned, private knowledge retrieval is based on private information retrieval. Chor et al.~\cite{ref:Chor.et.al.1995,ref:Chor.et.al.1998} proved that information-theoretical privacy can be achieved with the assumption that there are copies of the same database in at least two different queryable storage servers which do not communicate with each other. When this is the case, each server gets no information on the item retrieved by the user. In contrast, for a single server information-theoretic privacy can only be provided when the whole database is retrieved.

There are relaxations of this approach. For example, Kushilevitz and Ostrovsky~\cite{ref:Kushilevitz.et.al.1997} relaxed information-theoretical privacy to computational privacy and presented a method for constructing single-database PIR based on the Goldwasser-Micali public-key encryption scheme~\cite{ref:Goldwasser.et.al.1984}. Many other single-database PIR schemes have appeared since then. See the survey~\cite{ref:Ostrovsky.et.al.2007}, and more recently \cite{ref:Holzbaur.et.al.2020,ref:Bordage.et.al.2021,ref:Verma.et.al.2024} and the survey~\cite{ref:Alfarano.et.al.2023}.

In our implementation, we follow the approach of Chor et al.~\cite{ref:Chor.et.al.1995,ref:Chor.et.al.1998} to implement private knowledge retrieval using two copies of the database (see also~\cite{ref:Torra.2022:llibre}). For this, we need to index knowledge items. We have already specified above that the inference engine and the knowledge base clients request Horn clauses based on the pairs (predicate-name, arity). We use these pairs to index knowledge, and, therefore, we compute an index or hash from them. As the number of Horn clauses (facts and rules) and their lengths associated with a given index will be typically different, the database is padded so that all indices in the knowledge base have the same length.

To access a given index $i$, the agent client builds two random queries differing in the value associated to $i$. The agent submits one query to each knowledge base server, each server provides a string of bits that are xor-ed bits of the received query. These two responses are also xor-ed, and the resulting string is decoded providing the desired result, i.e., facts and rules associated to index $i$. 

The agent's knowledge-base client receives the requested knowledge, verifies its integrity using the associated provenance information and digital signature, caches it provided the verification is successful, and forwards it to the inference engine for inference. 

In addition to the approach that directly follows PIR, we have implemented two other alternative solutions that are more light-weighted. They are as follows. 

\begin{itemize}
\item Single server, $k$-anonymity~\cite{ref:Samarati.2001}. Given $k$, the agent submits the desired index $i$ and $k-1$ other random indices. The server provides all Horn clauses associated to the $k$ indices, and the client selects the desired one. 

\item Single server, buckets. All knowledge (Horn clauses) in the knowledge base is divided into $b$ buckets at the initialization stage. Then, for a desired index $i$, the agent determines in which bucket this knowledge is, and retrieves all associated knowledge (all Horn clauses) from the server. The agent then selects the desired one. 
\end{itemize}

In what follows, we will use the following notation. We denote by $n_{pa}$ the number of different pairs (predicate-name, arity) in the knowledge-base, by $n_h$ the number of Horn clauses in the knowledge base, and by $r$ the number of bits in the padded database. 

In summary, we have implemented three alternative private knowledge retrieval protocols. Each implementation has different formal privacy and security guarantees, as well as different computational requirements. The stronger privacy guarantees with the, in principle, largest computational time corresponds to the implementation of non-colluding knowledge bases. In contrast, $k$-anonymity and buckets are less computationally demanding but at the same time providing less privacy and security guarantees. Among $k$-anonymity and buckets, its privacy guarantees and computation requirements depend on the parameter $k$ and the number of buckets. So, if $b=n_{pa}/k$ then the privacy guarantees of the two approaches are similar, as a query will return a similar number of items. Note however that multiple access to the same index is deterministic for buckets but not for $k$-anonymity. 

\subsection{Untrusted agents and agent collusions}
In our implementation we have considered two servers to provide privacy preserving knowledge retrieval. In addition, knowledge is digitally signed to prevent tampering. Moreover, multiple servers provide interesting mechanisms for guaranteeing robustness and reliability for information storage \cite{ref:Shah.et.al.2014,ref:Tajeddine.et.al.2018}.
In this paradigm, encoding of the information stored by the servers (mainly by the use of maximum distance separable codes) prevents from the loss of information in case some servers get damaged or incommunicated and ensures reliability of the information, in case that a (small) number of servers send perturbed or directly fake information. This would provide additional security guarantees against memory poisoning attacks by some of the servers. However, when a number of servers is used, some of them may collude. Some solutions against this kind of collusions using error correcting codes can be found in \cite{ref:FreijHollanti.et.al.2017,ref:Holzbau.et.al.2022}.
In these references information retrieval is generalized to file downloading.
Another alternative is peer-to-peer PIR \cite{ref:Domingo.Ferrer.et.al.2009} in which a user is cloaked in a peer-to-peer user community, where peers submit queries on behalf of other peers and conversely.

Access to knowledge bases and ontologies can also be implemented using these approaches. In this way, untrusted collusions of servers will not know which knowledge items are accessed, and malicious actions could be mitigated. That is, we could minimize compromising memory and future agent's actions. Locally repairable codes (LRC)~\cite{{ref:Huang.et.al.2012},{ref:Sathiamoorthy.et.al.2013}}, that are used for large databases in practice, can be used for this purpose.

\section{Experiments and analysis}
\label{sec:4}
As described in the previous section, we have implemented a prototype of our architecture in Python. This prototype includes the local Prolog-like inference mechanism. Although all components run on a single laptop, we simulate both the knowledge-base server(s) and the knowledge-base client by providing separate software implementations for each role. 

 We are assuming that the server is not trusted, nevertheless, we do not allow severs to collude. The external knowledge base is indexed taking into account the predicate name and its arity. Therefore, the size of the database in number of records depends on how many distinctive pairs (name, arity) exist. Then, we have implemented a set of queries and evaluated their execution time experimentally. In addition, we assess the complexity of querying the system. We begin with the analysis of the complexity of the private information retrieval approaches introduced in Section~\ref{sec:implementation.PKR.3.casos}. 

\begin{table}[t]
\centering
\begin{tabular}{lcccc}
  \hline
  ~    & knowledge & sent          & received       & selection/  \\
  ~    & bases     & communication & communication  & computation$~~~$  \\
  \hline 
  $k$-anonymity &   1 KB & $k \cdot 64$         & $k \cdot r$  & select 1 of $k$ \\
  $b$-buckets   &   1 KB & $1 \cdot 64$         & $m \cdot r$  & select 1 of $b$ \\
  2-PIR       &   2 KB & $2 \cdot n_{pa}$      & $d r$          & XOR$(2 r)$      \\  
  $2^d$-PIR   & $d$ KB & $d \cdot n_{pa}^{1/d}$ & $d r$          & XOR$(d r)$      \\
  \hline
  \end{tabular}
\caption{Costs on bits for private knowledge retrieval implemented using $k$-anonymity, $b$-buckets, two non-colluding databases (2-PIR), and $2^d$ knowledge databases. 64 above represents the length in bits of integers (assuming long long int). Here, $r$ is the number of bits associated to each index in the knowledge base, $k$ the parameter associated to $k$-anonymity, $b$ is the number of buckets and $m=n_{pa}/b$ is the number of rows retrieved by $b$-buckets. \label{tab:costs.pkr}}
\end{table}

\subsection{Complexity analysis}
In the case of two non-colluding knowledge bases, each one with $n_{pa}$ keys, the agent is required to send $n_{pa}$ bits to each server (so, communication requires $2n_{pa}$ bits) and this agent will receive $r$ bits from each server (so, communication requires $2r$ bits). Here $r$ corresponds to the number of bits of a knowledge item of the padded database. Then, the agent performs a xor of these $2r$ bits obtaining a string of $r$ bits. Our solution is not optimal. There are alternative solutions for $2^d$ databases for a given $d$ in which the protocol sends $d\cdot n_{pa}^{1/d}$ bits in total and the agent receives from each database $r$ bits (i.e., $dr$ bits are received). Here the agent applies a xor of these $d \cdot r$ bits obtaining $r$ bits.

In the case of $k$-anonymity, the intruder sends $k$ indices, say integers of 64 bits (so, it sends $k \cdot 64$ bits) and receives $k$ knowledge items. Considering as above that they are $r$ bits long, this means that $k \cdot r$ bits are received. These strings need to be filtered to obtain the final $r$ bits.

In the case of buckets, the approach has a cost analysis similar to the one of $k$-anonymity, except for the transmission from client to server, which is just the number associated to a set of buckets. So, say, 64 bits. Then, for a number of buckets equal to $b$ we will have that each query will retrieve $m=n_{pa}/b$ indices from the knowledge base. So, the agent will receive $m \cdot r$ bits that need to be filtered to obtain the final $r$ bits. Costs are summarized in Table~\ref{tab:costs.pkr}.

\subsection{Computational experiments}
To test our system, we have defined a knowledge base consisting of $n_{h}=99$ facts and rules, which corresponds to $n_{pa}=46$ entries in the database. We have included a variety of Horn clauses, including both facts and rules, with non-recursive and recursive rules, and implementing as well two predicates that are benchmarks from SICStus Prolog~\cite{ref:sicstus.benchmarks}. In particular, we have implemented {\tt nreverse} (naive reverse) which for a list of $n=30$ elements requires a recursion of depth $(n+1)+2n(n+1)=31+465=496$, and {\tt zebra}. 

Then, we have considered a set of 18 queries and executed them in different configurations. We have also executed each configuration several times. The configurations we have considered are the following ones.

\begin{itemize}
\item ``baseline'' which means that a single external knowledge base is used, no privacy preserving access is applied, 
\item ``pir\_bucket'' which means that we use a single external knowledge base implementing private information retrieval with buckets,
\item ``pir\_kAnon'' which means that we use a single external knowledge base implementing private information retrieval with $k$-anonymity, and
\item ``itpir\_2servers'' which means that we use the PIR protocol with two external knowledge bases.
\end{itemize}

\begin{table}[t]
\centering
\begin{tabular}{lcccc}
\hline
\textbf{KB} & \textbf{Mean} & \textbf{St. dev.} & \textbf{Min} & \textbf{Max} \\
\hline
baseline & 0.17787843 & 0.00276265 & 0.17516215 & 0.18181177 \\
bucket   & 0.16892560 & 0.00250258 & 0.16619691 & 0.17300868 \\
Anon k4  & 0.17944365 & 0.00301458 & 0.17585521 & 0.18262578 \\
Anon k6  & 0.17852928 & 0.00138278 & 0.17638855 & 0.17977013 \\
Anon k8  & 0.18061575 & 0.00335783 & 0.17653908 & 0.18402709 \\
Anon k10 & 0.17887916 & 0.00275545 & 0.17525422 & 0.18126626 \\
it2pir   & 0.22309446 & 0.00500567 & 0.21711705 & 0.22866296 \\
\hline
\end{tabular}
\caption{Execution time statistics for different knowledge base access mechanisms, five runs, each accessing the knowledge bases as first time (i.e., case Non-cached). \label{tab:five.times.five.new.access}}
\end{table}

Note that each of these types of executions are implemented for each of the above configurations. As queries in the external knowledge base are cached by the knowledge client, we have repeated executions in different ways. Computation times for each type of execution are provided. 

\begin{itemize}
\item {\bf Non-cached.} 5 executions of each query as if the query access was new each time but allowing caching for recursion as well as within the set of queries. For example, the first inference associated to {\tt nreverse} requires an external knowledge base access to retrieve the associated Horn clauses, but no external access is required for recursive computation of {\tt nreverse}. Moreover, no external access is required for another call of {\tt nreverse} within the same set of queries. 
\item {\bf Standard caching.} Two executions of a query with cache. That is, we have executed each set of queries twice. Nevertheless, external access is only applied once, and all the other executions use cached Horn clauses. Because of that, even if the first execution required external access, the second execution never requires this external access, and, thus, the overall time for executing the set of queries takes much less time. 
\item {\bf All cached.} Baseline with no external query. In order to assess the execution costs using the external knowledge base, we also provide mean execution times when no external knowledge base is used. These results are computed using only cached Horn clauses. 
\end{itemize}

\begin{table}[t]
  \centering
  \begin{tabular}{lcccc}
\hline
\textbf{KB} & \textbf{Mean} & \textbf{St. dev.} & \textbf{Min} & \textbf{Max} \\
\hline
Baseline   & 0.0969 & 0.1119 & 0.0178 & 0.1761 \\
Bucket     & 0.0919 & 0.1081 & 0.0154 & 0.1684 \\
Anon k4    & 0.1006 & 0.1142 & 0.0198 & 0.1814 \\
Anon k6    & 0.0976 & 0.1143 & 0.0168 & 0.1785 \\
Anon k8    & 0.0978 & 0.1129 & 0.0180 & 0.1776 \\
Anon k10   & 0.0987 & 0.1114 & 0.0200 & 0.1775 \\
it2pir     & 0.1192 & 0.1503 & 0.0130 & 0.2255 \\
\hline
\end{tabular}
\caption{Execution time statistics for different knowledge base access mechanisms corresponding to two runs with cache. That is, first execution of the queries uses external access, and the second never does (case Standard caching). \label{tab:two.times.one.access}}
\end{table}

The execution times for the first set of experiments with the minimum number of cached Horn clauses (case Non-cached) are provided in Table~\ref{tab:five.times.five.new.access}. Then, execution times for the second set of experiments with partial caches (case Standard caching) are provided in Table~\ref{tab:two.times.one.access}. Finally, execution times for all Horn clauses cached (case All cached) are provided in Table~\ref{tab:five.times.all.cached}. As expected, the longest execution time is for the case non-cached, Table~\ref{tab:five.times.five.new.access}. 

The computational results displayed in these tables for the two PIR databases are surprisingly good. In the case non-cached, its time is 0.17s for baseline and 0.22s for it2pir. It seems that the implementation of the XOR on the knowledge base is extremely fast and comparable to the other expected-light-weight implementations that need to generate multiple queries, forward these queries, and then filter the results of these queries. Query generation and post-processing is implemented in Python.

Note that as all the query processing is performed by the knowledge base client and knowledge base server, execution times are in fact independent of the logical inference engine (our Prolog engine). In other words, the Prolog engine is agnostic on the type of knowledge base server used and it only queries the local knowledge base client and gets appropriate Horn clauses in return.

\begin{table}[t]
\centering
\begin{tabular}{lcccc}
\hline
\textbf{KB} & \textbf{Mean} & \textbf{St. dev.} & \textbf{Min} & \textbf{Max} \\
\hline
Baseline   & 0.0135 & 0.0027 & 0.0117 & 0.0176 \\
Bucket     & 0.0149 & 0.0038 & 0.0120 & 0.0205 \\
Anon k4    & 0.0158 & 0.0044 & 0.0121 & 0.0217 \\
Anon k6    & 0.0140 & 0.0016 & 0.0125 & 0.0161 \\
Anon k8    & 0.0126 & 0.0016 & 0.0116 & 0.0150 \\
Anon k10   & 0.0144 & 0.0021 & 0.0131 & 0.0175 \\
it2pir     & 0.0129 & 0.0007 & 0.0125 & 0.0140 \\
\hline
\end{tabular}
\caption{Execution time statistics for different KB access mechanisms corresponding to five runs with no external access (i.e., baseline, case All cached). \label{tab:five.times.all.cached}}
\end{table}

\section{Conclusions}
In this paper we claim that the standard formalization of memory poisoning attacks for semantic memory can be mitigated using secure memory mechanisms, secure updating of knowledge bases, and secure communication and communication logs. We implemented a knowledge base which includes provenance and digital signatures to avoid knowledge tampering. We introduced a private knowledge retrieval and private inference as a way to provide secure inference. We described our implementation and provided experiments and analysis of its cost. 

We mentioned in the paper that there are some ways to improve the performance of the inference system that were not considered and are left as possible extensions. In relation to knowledge retrieval, we considered retrieval by predicate name and arity. More sophisticated retrievals are possible taking into account the arguments of the predicates (e.g., as Just In Time Indexing, JITI, utilities in SWI-Prolog). Also, it would be possible to use tabling, memoization of inference results.

In our implementation we considered only a single external knowledge base, or two knowledge bases with replicated content. Our approach can be easily extended to multiple knowledge base servers with different knowledge. Specialized knowledge bases will offer their services listing their knowledge (i.e., providing appropriate hashing functions). In this case, agent's client would proceed in a way analogous to our current system, but deciding first to which server it needs to send a request. I.e., instead of having a single hash function providing an index for the pair predicate name/arity we would have multiple hash functions, one for each knowledge base.

\section*{Acknowledgements}
This work was partially supported by the Wallenberg AI, Autonomous Systems and Software Program (WASP) funded by the Knut and Alice Wallenberg Foundation. Support by Swedish Research Council (VR 2023-05531) is also acknowledged. The work was also partially supported by project PID2024-156636NB-C21 (MATSE) funded by MCIN/AEI/10.13039/501100011033/ FEDER, UE.

\end{document}